\documentclass[12pt,reqno]{amsart}

\usepackage{amsthm,amsmath,amsfonts,epsfig,lscape}

\usepackage{caption}
\usepackage{float}
\usepackage{placeins}
\usepackage{algorithm}
\usepackage{hyperref}

\usepackage{mathtools}
\usepackage{comment}
\usepackage{algorithmic}
\usepackage{color}

\theoremstyle{plain}

\theoremstyle{definition}

\theoremstyle{remark}

\newcommand{\defeq}{\mathrel{\mathop:}=}

\allowdisplaybreaks[4]

\begin{document}

\title[High-dimensional inference with the LOCO path]{High-Dimensional Inference Based on the Leave-One-Covariate-Out LASSO Path}
\author{Xiangyang Cao, Karl Gregory, and Dewei Wang}
\thanks{Department of Statistics, University of South Carolina, 216 LeConte College, 1523 Greene St, Columbia, SC, 29201, USA}

\date{}
\subjclass[2010]{Primary 62J07; secondary 62F40}
\keywords{high-dimensional inference, variable importance, variable selection, simultaneous inference, bootstrap}
\maketitle

\begin{abstract}
We propose a new measure of variable importance in high-dimensional regression based on the change in the LASSO solution path when one covariate is left out. The proposed procedure provides a novel way to calculate variable importance and conduct variable screening. In addition, our procedure allows for the construction of P-values for testing whether each coefficient is equal to zero as well as for testing hypotheses involving multiple regression coefficients simultaneously; bootstrap techniques are used to construct the null distribution. For low-dimensional linear models, our method can achieve higher power than the $t$-test. Extensive simulations are provided to show the effectiveness of our method. In the high-dimensional setting, our proposed solution path based test achieves greater power than some other recently developed high-dimensional inference methods. 
% Our method naturally extends to generalized linear models and can be applied to solution paths of other regularization methods. 
\end{abstract}

\section{Introduction}

We consider the linear regression model
\begin{equation}\label{eqn:linearmodel}
Y = \mathbf{X}\beta + \epsilon,
\end{equation}
where $\mathbf{X} = [X_1^T,X_2^T,\dots,X_n^T]^T$ with $X_i \in \mathbb{R}^p $, $Y \in \mathbb{R}^n$,  $\epsilon \sim \mathcal{N}(0,\sigma_{\epsilon}^2\,\mathbf{I}_n)$, where $\mathbf{I}_n$ is the $n\times n$ identity matrix, and $\beta\in\mathbb{R}^p$ is a vector of unknown regression coefficients.  We consider both the cases $p > n$ and $p \leq n$.
\par
We propose a measure of variable importance based on the change in the LASSO solution path due to removing a covariate from the model.  Regarding the LASSO solution path
\begin{equation}\label{eqn:lassopath}
    \hat{\beta} \defeq \hat{\beta}(\lambda) = \operatorname*{argmin}_{\beta \in \mathbb{R}^p} (||Y - \mathbf{X}\beta ||_{2}^{2} + \lambda||\beta||_{1} ),\quad \lambda > 0
\end{equation}
as a function of $\lambda$ taking values in $(0,\infty)$ and returning values $\hat{\beta}(\lambda)$ in $\mathbb{R}^p$, we propose to measure the importance of covariate $X_j$, for any $j\in\{1,\dots,p\}$, by comparing the path $\hat \beta$ to the path
\begin{equation}\label{eqn:LOCOpath}
    \hat{\beta}^{(-j)} \defeq \hat{\beta}^{(-j)}(\lambda) = \operatorname*{argmin}_{\beta \in \mathbb{R}^p, \beta_j=0} (||Y - \mathbf{X}\beta ||_{2}^{2} + \lambda||\beta||_{1} ),\quad \lambda > 0,
\end{equation}
which is the LASSO solution path when the covariate $X_j$ is removed from the model. Herein, for a vector $\mathbf{v} = (v_1, \dots, v_K)^T$, $||\mathbf{v}||^{2}_{2}=\sum_{k=1}^{K}v^2_k$ and  $||\mathbf{v}||_{1}=\sum_{k=1}^{K}|v_k|$. We will refer to $\hat \beta^{(-j)}$ as the leave-one-covariate-out solution path, or the LOCO path of the LASSO. It is important to note that for a given $j$, $\hat \beta^{(-j)}_j(\lambda)=0$ for all $\lambda$. We reason that if covariate $X_j$ is important, its importance will be reflected in a large difference between the paths $\hat \beta$ and $\hat \beta^{(-j)}$, whereas if it %covariate $X_j$ %
is not important, the difference between the paths $\hat \beta$ and $\hat \beta^{(-j)}$ will be small.  
\par 
The measure of variable importance we propose, which we shall call the LOCO path statistic, can be used for variable selection and variable screening; moreover, we suggest that it can be used as a test statistic for testing the hypotheses $H_0$: $\beta_j = 0$ versus $H_1$: $\beta_j \neq 0$. We also use the LOCO solution path idea to construct a test statistic for testing more complicated hypotheses involving several coefficients, specifically hypotheses of the form
\[
\text{$H_0$: $\beta_j = \beta_{j,0}$, for all $j \in \mathcal{A}$ versus $H_1$: $\beta_j \neq \beta_{j,0}$ for some $j \in \mathcal{A}$},
\]
for some $\{\beta_{j,0}, j \in \mathcal{A}\}$, where $\mathcal{A}\subset\{1,\dots,p\}$.  We propose a bootstrap procedure to calibrate the rejection regions of hypothesis tests based on the LOCO solution path.
\par
We now place our ideas in the literature: the LASSO was introduced in 1996 in \cite{tibshirani1996regression}, and has since been one of the most popular estimators for the linear regression model of \eqref{eqn:linearmodel}, particularly in the $p>n$ case.  It belongs to a class of penalized estimators designed to promote sparsity among the estimated regression coefficients in order to achieve simultaneous variable selection and estimation.  Implementing the LASSO requires choosing a value, usually via cross validation, of the tuning parameter $\lambda$, which governs the sparsity and shrinkage towards zero of the estimated regression coefficients.  Although the LASSO is a powerful tool, the LASSO estimator has a very complicated sampling distribution, so that statistical inference based on LASSO estimators is problematic.
\par
Other estimators for model \eqref{eqn:linearmodel} with $p>n$ have been proposed which have, under some conditions, limiting normal distributions, such as the desparsified LASSO estimator introduced by \cite{van2014asymptotically} and \cite{zhang2014confidence} as well as the estimator introduced by \cite{javanmard2014confidence}; these methods enable inference, but a downside is that they require the choice of an additional tuning parameter and inferences may be very sensitive to the choice of tuning parameter. The adaptive LASSO estimator of \cite{zou2006adaptive}, under some conditions and with tuning parameters appropriately chosen, has a limiting normal distribution (for non-zero coefficients), though convergence seems to be slow; a bootstrap procedure has been shown to be consistent for the adaptive LASSO in \cite{das2019perturbation}. A bootstrap method for the LASSO is proposed in \cite{chatterjee2011bootstrapping} and \cite{chatterjee2013rates}, which is consistent for a modified LASSO and adaptive LASSO estimator. A sequential significance testing procedure for variables entering the model along the LASSO solution path was proposed in \cite{lockhart2014significance}.  Inferential methods for the high-dimensional linear model based on sample splitting, for example in \cite{wasserman2009high} and \cite{meinshausen2009p}, have also been proposed and implemented with success.   

As variable selection methods, sure independence screening (SIS) and iterative sure independence screening (ISIS) are proposed in \cite{fan2008sure} for ultra-high dimensional linear regression. Ultra-high dimensional regression focuses on the settings with $\text{log}(p) = O(n^{\zeta})$. It has been extended to GLM \cite{fan2010sure}, GAM \cite{fan2011nonparametric} and multivariate regression models \cite{ke2014covariance}.  Although these methods enjoy the sure screening property \cite{fan2008sure}, SIS only considers the marginal contribution of each variable to the response.  

\par 
To our knowledge, however, not much work has focused on analyzing and summarizing the information contained in the entire solution path of the LASSO with respect to the importance of each variable.  We propose to consider the LASSO solution path in its entirety, and then measure how it changes when we leave one covariate out.
\par
The idea of leave-one-covariate-out (LOCO) inference is not new. The following LOCO-based procedure for measuring variable importance is described in \cite{lei2018distribution}:  Let $\hat{\mu}$ be an estimate of $E(Y|\mathbf{X})$ based on some training data $(\mathbf{X}, Y)$, and let $ \hat{\mu}_{(-j)} $ be the same estimator based on the training data $(\mathbf{X}_{(-j)}, Y)$, where $\mathbf{X}_{(-j)}$ is the matrix $\mathbf{X}$ with column $j$ removed. Then we measure the excess prediction error on new data $(X_{new}, Y_{new})$ as
\[
    |Y_{new} - \hat{\mu}_{(-j)}(X_{new})| - |Y_{new} - \hat{\mu}(X_{new})|,
\]
where the ``new'' data can come from crossvalidation testing sets or from a separate testing data set. The larger the above quantity, the greater importance we assign to covariate $X_j$, as it measures how much worse our predictions become due to removing covariate $X_j$.
\par
Permutation feature importance, introduced by \cite{breiman2001random} and generalized by \cite{fisher2018all}, is a similar to the LOCO approach to measuring variable importance; instead of removing covariate $X_j$ from the model, the observed values of covariate $X_j$ are randomly permuted. By this permutation, the association between covariate $X_j$ and the response is broken and the resulting model is different from the one fit to the original data.
\par

What we propose falls into the framework of LOCO variable importance and inference; however, rather than measuring the change in the prediction error due to removing a covariate, we consider the change in the LASSO solution path.

This paper is organized as follows: Section \ref{sec:LOCOpathstatistic} defines our measure of variable importance based on the change in the LASSO solution path due to the removal of a covariate and discusses its use as a variable selection and variable screening tool.  Section \ref{sec:hypothesistesting} explains how we propose to use the LOCO solution path idea to construct test statistics for testing hypotheses about the regression coefficients. We also describe a bootstrap procedure for estimating the null distribution of our LOCO path-based test statistics. Section \ref{sec:simulations} presents simulation results and Section \ref{sec:dataanalysis} illustrates the method on a real data set.  Section \ref{sec:discussion} provides additional discussion.

 %Although LASSO is popular, LASSO estimates are biased \cite{zou2006adaptive} and variable selection consistency depends on many unrealistic assumptions \cite{zhao2006model}. People proposed SCAD \cite{fan2001variable}, MCP \cite{zhang2010nearly} and Adaptive LASSO \cite{zou2006adaptive} to alleviate the bias of LASSO and make more consistent variable selection under fewer assumptions. However, these methods hinge on a tuning parameter $\lambda$, which needs to be carefully selected. Moreover, they cannot provide inference like p-value or confidence interval \cite{dezeure2015high}, because the distributions of regression coefficients follow some complicated non-normal distribution. 

\section{The leave-one-covariate-out path statistic}\label{sec:LOCOpathstatistic}

% If we apply squared error loss and the $l_1$ penalty, the estimator in \eqref{eqn:minimizationproblem} becomes the LASSO estimator

% \begin{equation}\label{eqn:lassopath}
%     \hat{\beta} = \hat{\beta}(\lambda) \defeq \operatorname*{argmin}_{\beta \in \mathbb{R}^p} (||Y - \mathbf{X}\beta ||_{2}^{2} + \lambda||\beta||_{1} )
% \end{equation}
% presented in \cite{tibshirani1997lasso}.
% \par

% Our method is based on a very intuitive idea, if we remove a truly significant variable, say, $\beta_j$ and obtain $\hat{\beta}^{(-j)}(\lambda)$ by solving

% \begin{equation}
%     \hat{\beta}^{(-j)} = \hat{\beta}^{(-j)}(\lambda) \defeq \operatorname*{argmin}_{\beta \in \mathbb{R}^p, \beta_j=0} (||Y - \mathbf{X}\beta ||_{2}^{2} + \lambda||\beta||_{1} ),
% \end{equation}
% then the difference between the solution paths $\hat{\beta}(\cdot)$ and $\hat{\beta}^{(-j)}(\cdot)$ should be large, or at least larger than if we had removed an insignificant variable. Hence the difference in the LASSO solution path due to removing a covariate could be viewed as a measure of importance for the covariate. 
% \par
To formulate our metric for the difference between the LASSO solution path $\hat \beta$ defined in \eqref{eqn:lassopath} and the LOCO solution path of the LASSO defined in \eqref{eqn:LOCOpath}, we define a quantity for functions taking values in $(0,\infty)$ and returning values in $\mathbb{R}^p$. Firstly, for any function $g$ taking values in $(0,\infty)$ and returning values in $\mathbb{R}$, let
\[
\|g\|_s = \begin{dcases}
            (\hbox{\text{$\int_0^\infty |g(\lambda)|^s d\lambda$}})^{1/s},& 0 < s < \infty\\
            \sup_{\lambda > 0}|g(\lambda)|, & s = \infty.
\end{dcases}
\]
Secondly, for a vector $x \in \mathbb{R}^p$, let
\[
||x||_t  = \begin{dcases*}
        (\hbox{\text{$\sum_{j=1}^p|x_j|^t$}})^{1/t}, & $0 < t < \infty$ \\
        \operatorname*{max}_{1 \leq j \leq p}|x_j|,& $ t = \infty$.
        \end{dcases*}
\]
Now, for a function $f$ taking values in $(0,\infty)$ and returning values in $\mathbb{R}^p$ such that $f(\lambda) = (f_1(\lambda),\dots,f_p(\lambda))^T$, define the quantity $\|f\|_{s,t}$ as
\[
\|f\|_{s,t} = \|(\|f_1\|_s,\dots,\|f_p\|_s)^T\|_t.
\]
Having defined a quantity for functions taking values in $(0,\infty)$ and returning values in $\mathbb{R}^p$, we define the LOCO path statistic for covariate $X_j$ as
$$
T_j(s,t) = || \hat \beta - \hat \beta^{(-j)} ||_{s,t},
$$
which measures the change in the LASSO solution path due to removing covariate $X_j$ from the model.
\par
In practice, it is convenient to use $s = t$; if $s = t = q$, we have 
$$
    T_j(q,q) = \begin{dcases*}
    \left (\sum_{k=1}^p\int_0^{\infty} |\hat{\beta}_k(\lambda) - \hat{\beta}_k^{(-j)}(\lambda)|^q d\lambda \right)^{\frac{1}{q}} &  $q < \infty$ \\
    \max_{1\leq k \leq p} \sup_{\lambda > 0} |\hat{\beta}_k(\lambda) - \hat{\beta}_k^{(-j)}(\lambda)| &  $q = \infty$. \\
    \end{dcases*}
$$
We recommend using $q=1$ or $q=2$ in practice. We have found that under $q=\infty$ our hypothesis test tend to have lower power, so we do not recommend this setting. We illustrate this in the simulation section.

We posit that the quantity $T_j(s,t)$ will be large if $\beta_j \neq 0$ and small if $\beta_j = 0$, for $j=1,\dots,p$, so that $T_j(s,t)$ may serve as a measure of variable importance for covariate $X_j$. Since the LASSO solution path is piecewise linear, we can calculate $T_j(s,t)$ exactly. More details about the calculation can be found in the section S.1 of the Supplementary Material.

\subsection{The LOCO path statistic as a measure of variable importance}

% \rd{ Most common method for measuring variable importance utilized the following idea: by ruling out the variable of interest, we measure how this would change our model. Different methods may use different ways of ruling out or different metrics for measuring the change. The most common approach is measuring the prediction performance after ruling out. }

\begin{figure}[htbp]
  \includegraphics[scale = 0.2]{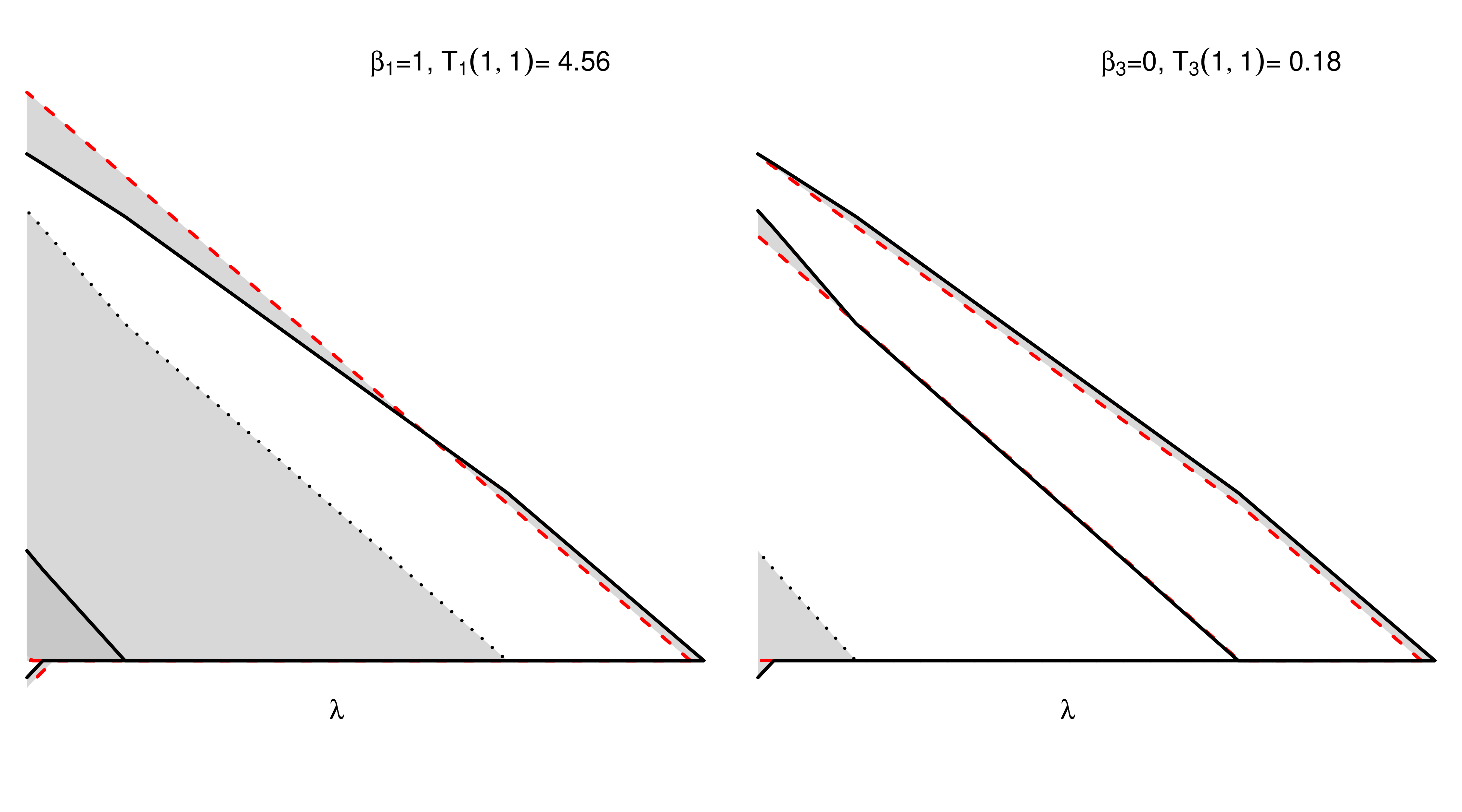}
  \centering
  \caption{Shaded areas show how $T_j(1,1)$ measures the change in LASSO path. Black solid line depicts the solution path before removal. Black dotted line depicts the solution path of the covariates being removed. Red dashed line depicts the solution path after removal. Left: $T_1(1,1)$. Right: $T_3(1,1)$.}
  \label{fig:var_imp_two}
\end{figure}

% \rd{Our method, in these cases, provide a new perspective on these metrics. We won't calculate the change in prediction accuracy, since this requires tuning and sample splitting. We measure the change in the regularization path. Notice this idea is not limited to the LASSO on linear model. Any regularized machine learning methods are potential candidate for our method.
% }
For the sake of illustration, let us consider one special case of $T_j(s, t)$, with $s=t=1$. We have

\begin{equation*}
    T_j(1,1) = ||\hat{\beta} - \hat{\beta}^{(-j)}||_{1,1} = \sum_{k=1}^p\int_0^{\infty} |\hat{\beta}_k(\lambda) - \hat{\beta}_k^{(-j)}(\lambda)| d\lambda,
\end{equation*}

which is equal to the sum of all the areas under the curves $|\hat{\beta}_k(\cdot) - \hat{\beta}_k^{(-j)}(\cdot)|$, $k=1,\dots,p$. We depict this for the following simple example: We generate one dataset from the linear regression model \eqref{eqn:linearmodel} with $n = 100$, $p = 4$ and $\beta = (1,1,0,0)^T$, and compute the test statistics $T_1(1,1)$ and $T_3{(1,1)}$. The left and right panels of Figure \ref{fig:var_imp_two} show the original LASSO solution path as well as the solution path after removing the first and third covariates, respectively, from the model. In each panel, the sum of the areas of the shaded regions is the value of the test statistic.

%Figure \ref{fig:var_imp_all} depicts the calculation of the values $T_1(1,1),\dots,T_{12}(1,1)$ on the same simulated data set.  We see that when covariates with larger coefficients are removed from the model, the change in the solution path is greater than when covariates with smaller or zero-valued coefficients are removed from the model.

% As we expected, if we remove a variable with larger $\beta_i$, our method assign a larger value on that variable. And from Figure 1 and 2, we also observed a larger shaded area. 

%\begin{figure}[htbp]
%  \includegraphics[scale = 0.5]{var_imp_all.pdf}
%  \centering
%  \caption{Variable importance for all variables.}
%  \label{fig:var_imp_all}
%\end{figure}

\begin{figure}[htbp]
  \includegraphics[scale = 0.5]{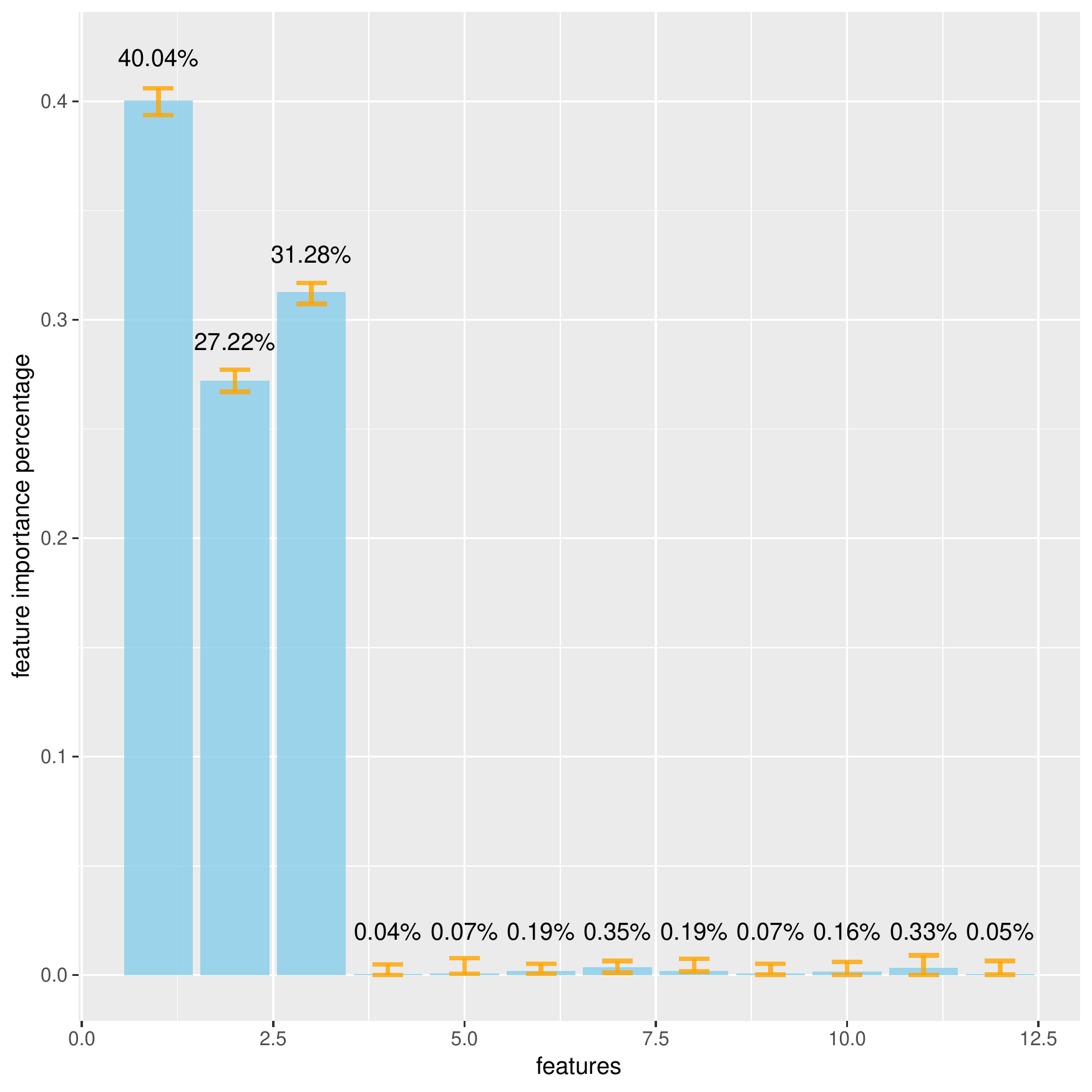}
  \centering
  \caption{\small Variable importance for all variables based on the LOCO path statistic. The error bar is our permutation interval. The variable importance is also shown as percentages on top of the error bar.}
  \label{fig:var_imp_bar}
\end{figure}

We propose to summarize the importance of the variables measured by the LOCO path statistic in the following way. After standardizing the values of $T_{j}(s,t)$, $j=1,\dots,p$, so that they sum to one, for example by defining
\[
\overline T_j(s,t) = T_j(s,t)\left(\sum_{k=1}^p T_k(s,t)\right)^{-1}, \quad j = 1,\dots,p,
\]

we can make a plot such as the one in Figure \ref{fig:var_imp_bar}, which shows the values of $\overline T_1(1,1),\dots,\overline T_{12}(1,1)$, expressed as percentages. This is based on a single dataset simulated from \eqref{eqn:linearmodel} with $n = 100$, $p = 12$, 
$\beta = (1,1,1,0,\dots,0)^T$, for the sake of illustration. %$\mathbf{X} \sim \operatorname{MVN}(\mathbf{0}, \mathbf{I}_n)$,
%and $\epsilon \sim \mathcal{N}(0, \mathbf{I}_n)$. 
The first three covariates are seen to have the highest importance according to the LOCO path statistic.
\par
Furthermore, we consider attaching to the variable importance a measure of uncertainty. The LOCO path $\hat{\beta}_k^{(-j)}(\lambda)$ could be fitted by permuting variable $j$ in $\mathbf{X}$. By permuting variable $j$ in $\mathbf{X}$, we break the association between $X_j$ and $Y$, which has an effect similar to removing variable $j$. By permuting the observed values of covariate $j$ multiple times we can obtain an interval for the variable importance. Figure \ref{fig:var_imp_bar} also shows the permutation interval calculated for the importance measure of each variable.

\subsection{Variable screening in ultra-high dimensional settings}

The so-called ultra-high dimensional setting was discussed in \cite{fan2008sure}, where the dimensionality $p$ grows exponentially ($\text{log}(p)=O(n^{\zeta})$) as $n$ grows. For ultra-high dimensional problems, preliminary variable screening is often done to reduce the dimension of the data.
\par

Our method naturally adapts to ultra-high dimensional settings. By calculating how the removal of each variable will alter the LASSO solution path, we have a simple way to screen out variables which are likely to be irrelevant. Our method uses the information contained in the LASSO solution path, which utilizes both joint and marginal information. One interesting result of LASSO in the high-dimensional setting is that some variables never enter the model. %
If we take a closer look at the solution path of such variables, they are equal to $0$ for all values of $\lambda$. If we were to use cross validation to select the LASSO tuning parameter and obtain the final selection results, these variables would never be selected. This means we can safely screen out these variables at the beginning. 
\par

Based on this intuition, we suggest the following screening procedure: Compute the solution path with all variables in the model. Then remove one variable at a time and compute the LOCO solution path; compute the values $T_{1}(s,t),\dots,T_p(s,t)$, which compare the solution path based on the full set of covariates to the LOCO solution paths.  Then screen out variables for which $T_j(s,t) \leq \epsilon$, where $\epsilon$ is a user-specified threshold. Choosing $\epsilon=0$ discards only those variables which never enter the solution path.  We can also rank $T_j(s,t)$ and only select the top $K$ variables, where we might choose $K$ to be $n-1$ and $n$ is the sample size.
% \begin{enumerate}

% \item Train the model and fit the whole regularization path.

% \item For $j \in \{1,2,\dots, p\}$, removed one variable $j$, and fit the regularization path again.

% \item Calculate $T_j(s,t)$.

% \item Repeat the 2nd and 3rd step until we exhaust all variables.

% \item Screen out variables with $T_j(s,t) < \epsilon$. 

% \end{enumerate}

%We find that when $p$ is very large, it can be very time-consuming to compute the entire LASSO solution path even once---and the above procedure requires computing $p+1$ solution paths, one path with all the covariates and each of the $p$ LOCO paths. To speed up the variable screening procedure, we propose, rather than fitting the entire solution path (for example using the LASSO-modification to the LARS algorithm), we can fit the solution paths over a coarse grid of $\lambda$ values; this will result in some loss of precision in the computation of $T_j(s,t)$, but will dramatically reduce the computation time. Moreover, for variable screening, since it is only important whether $T_j(s,t) < \epsilon$, we believe we can afford to lose some precision at this step. 

% In other words, by screening out variables with zero or close-to-zero variable importance, we aim to reduce the dimension of the data to a scale on which . Hence we can continue to use other method.

\section{Hypothesis testing using the LOCO path idea}\label{sec:hypothesistesting}

We now consider using the LOCO path idea to test hypotheses of the form
\begin{equation}\label{eqn:simultaneousH0}
\text{$H_0$: $\beta_j = \beta_{j,0}$ for all $j \in \mathcal{A}$ versus $H_1$: $\beta_j \neq \beta_{j,0}$ for some $j \in \mathcal{A}$},
\end{equation}
for some $\{\beta_{j,0}, j \in \mathcal{A}\}$, where $\mathcal{A}\subset\{1,\dots,p\}$.  We first calculate the LASSO solution path with all variables included.  Next, we compute the solution path subject to the constraint specified by the null hypothesis, which is given by
\begin{equation}\label{eqn:solpathH0}
    \hat{\beta}_0 \defeq \hat{\beta}_0(\lambda) = \operatorname*{argmin}_{\beta \in \mathbb{R}^p, \beta_j = 0 \in \mathcal{A}} (||(Y - \mathbf{X}_\mathcal{A}\beta_{0,\mathcal{A}}) - \mathbf{X}\beta ||_{2}^{2} + \lambda||\beta||_{1} ),
\end{equation}
where $\beta_{0,\mathcal{A}} = (\beta_{j,0}, j\in\mathcal{A})^T$ and $\mathbf{X}_\mathcal{A}$ is the matrix constructed out of the columns of $\mathbf{X}$ with indices in $\mathcal{A}$.

% Testing  $H_0$: $\beta_j = 0$ versus $H_1$: $\beta_j \neq 0$, $ \hat{\beta}^{(-j)}(\lambda)$ still remains a popular research area. As we mentioned before, $\hat{\beta}^{(-j)}(\lambda)$ could be viewed as sort of a likelihood if we impose the constraint $H_0$: $\beta_j = 0$ on our model.
% In addition, we may want to test $H_0$: $\beta_j = \beta_0$ versus $H_1$: $\beta_j \neq \beta_0$ with $\beta_0 \neq 0$ sometimes, or a even more complicated hypothesis $H_0$: $f(\beta_j) = v$ versus $H_1$: $f(\beta_j) \neq v$. 

% Currently, we don't have a systematic way to solve all these type of hypothesis. However, by utilizing the idea of solution path, we can do the following. 

% We first calculate
% \begin{equation}
%     \hat{\beta}^{H_0} = \hat{\beta}^{H_0}(\lambda) \defeq \operatorname*{argmin}_{\beta \in \mathbb{R}^p, H_0: f(\beta_j) = v} (||Y - \mathbf{X}\beta ||_{2}^{2} + \lambda||\beta||_{1} )
% \end{equation}

We then suggest as a test statistic for testing $H_0$ versus $H_1$ the quantity 
\begin{equation}\label{eqn:teststatistic}
T_0(s,t) = \|\hat \beta - \hat \beta_0\|_{s,t},
\end{equation}
which compares the solution paths $\hat{\beta}_0$ and $\hat{\beta}$. For testing the hypotheses
\[
\text{$H_0$: $\beta_j = 0$ versus $H_1$: $\beta_j \neq 0$},
\]
for some $j\in\{1,\dots,p\}$, we have $\hat \beta_0 = \hat \beta^{(-j)}$, so that the test statistic $T_0(s,t)$ is equal to the LOCO path variable importance statistic $T_j(s,t)= \|\hat \beta - \hat \beta^{(-j)}\|_{s,t}$. 
\par

% For very complicated $f(\beta_j)$, we may need to design a new algorithm to solve (10). But for simple test like $H_0$: $\beta_j = \beta_0$, we can easily solve (10) by doing some transformation on $Y$. 

% Simultaneous inference focus on testing multiple hypothesis $H_{0,j}$: $\beta_j = 0$ versus $H_{1,j}$: $\beta_j \neq 0$ where $j \in \mathcal{A} $ and at the same time. Our method could be extended easily via 
% calculating the difference between $\hat{\beta}^{ \{H_{0,j}\} }(\lambda)$ and $\hat{\beta}(\lambda)$.
% \begin{equation}
%     \hat{\beta}^{ \{H_{0,j}\} } = \hat{\beta}^{\{H_{0,j}\}}(\lambda) \defeq \operatorname*{argmin}_{\beta \in \mathbb{R}^p, H_{0,j}, j \in \mathcal{A} } (||Y - \mathbf{X}\beta ||_{2}^{2} + \lambda||\beta||_{1} )
% \end{equation}

\subsection{A bootstrap estimator of the null distribution}

% To make statistical inference on $\beta$, we need to understand the null distribution of the test statistics. We propose a bootstrap approach for estimating the null distribution of our test statistics.
\par
In order to test the hypotheses in \eqref{eqn:simultaneousH0} using the test statistic $T_0(s,t)$ in \eqref{eqn:teststatistic}, we need to know the distribution of $T_0(s,t)$ under $H_0$.  We propose estimating this null distribution using a  residual bootstrap procedure.
\par
In order to obtain residuals from which to resample, we propose obtaining an initial estimator $\tilde \beta$, which we will discuss at the end of this section, of the vector $\beta$ from which we can obtain residuals
\[
\tilde \epsilon = Y - \mathbf{X} \tilde \beta.
\]
Let $\tilde Y^*$ be the $n \times 1$ random vector with entries given by $\tilde Y_i^* = X_i^T\tilde \beta + \tilde \epsilon_i^*$, for $i=1,\dots,n$, where $\epsilon_1^*,\dots,\epsilon_n^*$ are sampled with replacement from the entries of the residual vector $\tilde \epsilon = (\tilde \epsilon_1,\dots,\tilde \epsilon_n)^T$.
\par
% We then construct a bootstrap realization $\hat \beta^*$ of $\hat \beta$, imposing the null hypothesis $H_0$: $\beta_j = \beta_{j,0}$ for all $j \in \mathcal{A}$, as 
For testing the hypotheses in \eqref{eqn:simultaneousH0}, the bootstrap versions $\hat{\beta}^*$ and $\hat{\beta}_0^*$ of $\hat \beta$ and $\hat \beta_0$ are constructed as
\begin{equation}\label{eqn:bootstrapbeta}
    \hat{\beta}^* \defeq \hat{\beta}^*(\lambda) = \operatorname*{argmin}_{\beta \in \mathbb{R}^p} (||(\tilde Y^* - \mathbf{X}_\mathcal{A}(\tilde \beta_\mathcal{A} + \beta_{0,\mathcal{A}})) - \mathbf{X}\beta ||_{2}^{2} + \lambda||\beta||_{1} )
\end{equation}
and
\begin{equation}\label{eqn:bootstrapbetaH0}
    \hat{\beta}_0^* \defeq \hat{\beta}_0^*(\lambda) = \operatorname*{argmin}_{\beta \in \mathbb{R}^p, \beta_j = 0,j \in \mathcal{A}} (||(\tilde Y^* - \mathbf{X}_\mathcal{A}(\tilde \beta_\mathcal{A} + \beta_{0,\mathcal{A}})) - \mathbf{X}\beta ||_{2}^{2} + \lambda||\beta||_{1} ),
\end{equation}
respectively.  Then the bootstrap version of $T_0(s,t) = \|\hat \beta - \hat \beta_0\|_{s,t}$ is given by
\[
T_0^*(s,t) = \|\hat \beta^* - \hat \beta^*_0 \|_{s,t}.
\]
\par
Given a large number $B$ of Monte-Carlo replicates of $T_0^*(s,t)$, denoted by, say, $T_0^{*,(1)}(s,t)<\dots<T_0^{*,(B)}(s,t)$, when ordered, our bootstrap-based test of $H_0$ at significance level $\alpha$ has decision rule
\[
\text{Reject $H_0$ if and only if $T_0(s,t) > T_0^{*,(\lfloor B(1-\alpha)\rfloor)}$,}
\]
where $T_0^{*,(\lfloor B(1-\alpha)\rfloor)}$ is the Monte-Carlo approximation to the bootstrap estimator of the upper $\alpha$-quantile of the null distribution of $T_0(s,t)$, and $\lfloor \cdot \rfloor $ is the floor function. We could also obtain a bootstrapped P-value by 
\[
B^{-1}\sum_{i=1}^{B} I\{T_0^{*,(i)}(s,t) > T_0(s,t)\},
\]where $I(\cdot)$ is the indicator function.

\par
For the simpler hypotheses $H_0$: $\beta_j =0 $ versus $H_1$: $\beta_j \neq 0$ for any $j=1,\dots,p$, we need to construct a bootstrap version of the LOCO path statistic $T_j(s,t) = \|\hat \beta - \hat \beta^{(-j)}\|_{s,t}$.  The bootstrap versions of $\hat \beta$ and $\hat \beta^{(-j)}$, following \eqref{eqn:bootstrapbeta} and \eqref{eqn:bootstrapbetaH0}, are 
$$
    \hat{\beta}^* \defeq \hat{\beta}^*(\lambda) = \operatorname*{argmin}_{\beta \in \mathbb{R}^p} (||(\tilde Y^* - \mathbf{X}_j\tilde \beta_j) - \mathbf{X}\beta ||_{2}^{2} + \lambda||\beta||_{1} )
$$
and 
$$
    \hat{\beta}^{*(-j)} \defeq \hat{\beta}^{*(-j)}(\lambda) = \operatorname*{argmin}_{\beta \in \mathbb{R}^p, \beta_j = 0} (||(\tilde Y^* - \mathbf{X}_j\tilde \beta_j) - \mathbf{X}\beta ||_{2}^{2} + \lambda||\beta||_{1} ),
$$
respectively, where $\mathbf{X}_j$ is column $j$ of the matrix $\mathbf{X}$. Then the bootstrap version of $T_j(s,t)$ is given by
\[
T^*_j(s,t) = \|\hat \beta^* - \hat \beta^{*(-j)}\|_{s,t}.
\]

\par
Regarding the choice of the initial estimator $\tilde \beta$ of $\beta$, which is used only to obtain residuals suitable for resampling, we suggest, when $p \geq n$, the adaptive LASSO estimator 
$$
    \hat{\beta}^{Ada} = \operatorname*{argmin}_{\beta \in \mathbb{R}^p} (||Y - \mathbf{X}\beta ||_{2}^{2}+\gamma\sum^{p}_{j=1}\hat{w}_j|\beta_{j}| ),
$$
where the tuning parameter $\gamma$ is selected via 10-fold cross validation and the weights $\hat w_1,\dots,\hat w_p$ are given by
\[
\hat w_j = 1/|\hat \beta_j^L|, \quad j=1,\dots,p,
\]
where $\hat \beta_1^L,\dots,\hat \beta_p^L$ are the LASSO estimates of $\beta_1,\dots,\beta_p$ from \eqref{eqn:lassopath} under the 10-fold cross validation choice of $\lambda$. This is the initial estimator we have used in our simulation studies, and it appears to work well.  For the $p < n$ case the least-squares estimator could be used, though even in the low-dimensional case, we still recommend using the adaptive LASSO estimator when $p$ is close to $n$.

\subsection{Justification of the bootstrap for a simple case}

Finding the sampling distribution of $T_0(s,t)$ in general is a very hard problem which we do not attempt to solve. However, we do provide in this section an argument for why the bootstrap method described in the previous section will work in a simple case: the low-dimensional case, with $p<n$, with a design matrix having orthonormal columns. We focus on the null distribution of the test statistic $T_j(1,1) = \|\hat \beta - \hat \beta^{(-j)}\|_{1,1}$ for testing $H_0$: $\beta_j = 0$ versus $H_1$: $\beta_j \neq 0$ for some $j \in \{1,\dots,p\}$.
\par
In low-dimension, if the design matrix $\mathbf{X}$ satisfies $\mathbf{X}^T\mathbf{X} = \mathbf{I}_p$, where $\mathbf{I}_n$ is the $n \times n$ identity matrix, the LASSO solution path $\hat \beta$ has entries given by
\[
 \hat{\beta}_k(\lambda) = S_{\lambda}(\hat{\beta}_k^\text{LS}), \quad k = 1,\dots,p,
\]
where $\hat{\beta}^\text{LS}= (\mathbf{X}^T\mathbf{X})^{-1}\mathbf{X}^TY = \mathbf{X}^TY$ is the least-squares estimator of $\beta$ and $S_\lambda(\cdot)$ is the soft-thresholding operator defined by
\[
S_\lambda(x) = \begin{dcases}
x - \lambda, & x > \lambda \\
0, & - \lambda < x  <  \lambda \\
x + \lambda, & x < - \lambda
\end{dcases}
\]
for $\lambda \geq 0$. The solution path $\hat \beta^{(-j)}$ has entries given by
\[
 \hat \beta^{(-j)}_k(\lambda) = \begin{dcases}
 0 & k = j\\
 S_{\lambda}(\hat{\beta}_k^\text{LS}) & k \neq j\\
    \end{dcases}
\]
for $k=1,\dots,p$.
\par
In this case, the LOCO path statistic $T_j(1,1)$ is given by
\begin{align*}
    T_j(1,1) &=\| \hat \beta - \hat \beta^{(-j)}\|_{1,1}=\sum_{k=1}^p\int_0^{\infty} | \hat{\beta}_k(\lambda) - \hat{\beta}_k^{(-j)}(\lambda)| d\lambda\\
    &= \int_0^{|\hat{\beta}_j^{\text{LS}}|}(|\hat{\beta}_j^{\text{LS}}|-\lambda) d\lambda=\frac{1}{2}{|\hat{\beta}_j^{\text{LS}}|^{2}}.
\end{align*}
So, our test statistic is merely a $1$-to-$1$ mapping of the least-squares estimator. 
Hence, under $H_0$: $\beta_j = 0$, 
\begin{equation*}
nT_j(1,1) = \frac{n}{2}|\hat{\beta}_j^{\text{LS}}|^{2} \sim W \frac{\sigma^2}{2},
\end{equation*}
where $W \sim \chi_1^2$.  
\par
Now consider the bootstrap version $T^*_j(1,1)$ of $T_j(1,1)$ in the $p<n$ and orthonormal design case; we assume that the least-squares estimator is used as the initial estimator from which the residuals are obtained. Let $\hat \beta^{*,\text{LS}} = \mathbf{X}^T\tilde Y^*$ be the bootstrap version of $\hat \beta^{\text{LS}}$. Now, we can write the entries of
\[
\hat \beta^{*}(\lambda) = \operatorname*{argmin}_{\beta \in \mathbb{R}^p} (||(\tilde Y^* - \mathbf{X}_j\hat \beta_j^{\text{LS}}) - \mathbf{X}\beta ||_{2}^{2} + \lambda||\beta||_{1} )
\]
as
\[
\hat \beta_k^{*}(\lambda) = \begin{dcases}
        S_\lambda(\hat \beta_k^{*,\text{LS}} - \hat \beta_k^{\text{LS}}), & k=j\\
        S_\lambda(\hat \beta_{k}^{*,\text{LS}}), & k \neq j
        \end{dcases} \quad \text{ for } k =1,\dots,p,
\]
using the fact that
\[
\mathbf{X}^T_k(\tilde Y^* - \mathbf{X}_j\hat \beta_j^{\text{LS}}) = \begin{dcases}
\hat \beta_k^{*,\text{LS}} - \hat \beta_k^{\text{LS}}, &  k = j \\
\hat \beta_k^{*,\text{LS}} & k \neq j,
\end{dcases}\quad \text{ for } k = 1,\dots,p.
\]
In addition, we can write the entries of 
\[
\hat \beta^{*(-j)}(\lambda) = \operatorname*{argmin}_{\beta \in \mathbb{R}^p, \beta_j = 0} (||(\tilde Y^* - \mathbf{X}_j\hat \beta_j^{\text{LS}}) - \mathbf{X}\beta ||_{2}^{2} + \lambda||\beta||_{1} )
\]
as
\[
\hat \beta_{k}^{*(-j)}(\lambda) = \begin{dcases}
        0, & k=j\\
        S_\lambda(\hat \beta_{k}^{*,\text{LS}}), & k \neq j
        \end{dcases} \quad \text{ for } k =1,\dots,p.
\]
So we have
\begin{align*}
    T_j^*(1,1) &=\|\hat \beta^* - \hat \beta^{*(-j)}\|_{1,1}=\sum_{k=1}^p\int_0^{\infty} |\hat{\beta}_{k}^{*(-j)}(\lambda) - \hat{\beta}_k^{*}(\lambda)| d\lambda \\
    &= \int_0^{|\hat \beta_k^{*,\text{LS}} - \hat \beta_k^{\text{LS}}|}(|\hat \beta_k^{*,\text{LS}} - \hat \beta_k^{\text{LS}}|-\lambda) d\lambda=\frac{1}{2}{|\hat \beta_j^{*,\text{LS}} - \hat \beta_j^{\text{LS}}|^{2}}.
\end{align*}

It can be established that
\[
\operatorname*{sup}_{x \in \mathbb{R}} \left|P_*\left(\frac{n}{2}|\hat{\beta}^{*,\text{LS}}_{j} - \hat \beta^{\text{LS}}_j|^2
< x \right) - P\left(\frac{n}{2}|\hat{\beta}_j^\text{LS} - {\beta}_j|^2
 <x\right) \right| \xrightarrow{p} 0,
\]
as $n \to \infty$, where $P_*$ denotes probability conditional on the observed data \cite{mammen2012does}.  This means our bootstrap works in the low-dimensional orthonormal design case. In the high-dimensional case, or even in the low-dimensional case without the assumption of an orthogonal design, \eqref{eqn:lassopath} does not admit a simple solution, and in this setting the derivation of the distribution of the test statistic would be very difficult. Our simulation studies, however, suggest that our bootstrap procedure can consistently estimate the null distributions of the test statistics in the non-orthogonal design and high-dimensional cases.

\section{Simulation studies}\label{sec:simulations}

We now study via simulation the effectiveness of the LOCO path statistic as a variable screening tool as well as the properties of our proposed LOCO-path-based tests of hypotheses which use the residual bootstrap to estimate the null distributions of the test statistics. An R package \texttt{LOCOpath} that implements all of our proposed methods is publicly available at \url{http://github.com/devcao/LOCOpath}. We first present the variable screening results.

\subsection{Variable screening}

To assess the performance of the LOCO-path-based variable screening procedure described in Section \ref{sec:LOCOpathstatistic}, we follow the simulation example in  \cite{fan2008sure}, generating data from the model
 \[
Y = \beta X_1 + \beta X_2 + \beta X_3 + \epsilon,
\]
where $\epsilon \sim \mathcal{N}(0, 1)$, with a total of $p$ predictors $X_1, \dots, X_p$ in the model. The rows of the design matrix are generated as independent multivariate normal random vectors with covariance matrix $\Sigma = (\rho^{|i-j|})_{1\leq i , j \leq p}$, where $\rho = 0, 0.1, 0.5$ and $ 0.9$. Models with $\beta = 1, 2, 3$, $p = 100$, $n = 20$, and $p = 1000$, $n = 50$ are considered. We simulated 200 data sets for each model. To compare with SIS and ISIS, we utilized the R package \texttt{SIS} \cite{SIS_r}. We simulated 200 data sets and for each model we calculate $T_j(1,1)$ and $T_j(2,2)$ for $j = 1,2,\dots,p$ and select the top $n-1$ covariates, selecting the same number of covariates with SIS and ISIS in order to make a fair comparison. For our method, we utilized the R package \texttt{lars} \cite{lars_r} with LASSO modification to calculate our test statistic. 

In Table \ref{tab:var_screening_new} we show the proportion of times that the true model is contained in the set of selected covariates for our method and for the SIS and ISIS variable screening methods. In most cases, the model selected by our LOCO-path-based method contains the true model with greater frequency than that of the SIS and ISIS methods. We note that our method achieves this without any need for selecting tuning parameters, whereas the ISIS methods involves iterated LASSO fits for which the strength of the sparsity penalty must be chosen.

\begin{table}[htbp]
\centering
\begin{tabular}{llllll}
  \hline
  \hline
 Setting & $\beta$ & $T_j(1,1)$ & $T_j(2,2)$ & SIS & ISIS\\
  \hline
  \hline
  $p = 1000, n = 50, \Sigma = \mathbf{I}_p$ & 1 & 0.995 & 0.995 & 0.900 & 0.945 \\
  & 2 & 1 & 1 & 0.945 & 1 \\
  & 3 & 1 & 1 & 0.990 & 1 \\
  \hline
  $p = 1000, n = 50, \Sigma = (0.1^{|i-j|})_{1\leq i , j \leq p}$ & 1 & 0.990 & 0.990 & 0.960 & 0.960 \\
  & 2 & 1 & 1 & 0.995 & 1 \\
  & 3 & 1 & 1 & 0.990 & 1 \\
  \hline
 $p = 1000, n = 50, \Sigma = (0.5^{|i-j|})_{1\leq i , j \leq p}$ & 1 & 1 & 1 & 1 & 0.890 \\
  & 2 & 1 & 1 & 1 & 1 \\
  & 3 & 1 & 1 & 1 & 1 \\
    \hline
 $p = 1000, n = 50, \Sigma = (0.9^{|i-j|})_{1\leq i , j \leq p}$ & 1 & 0.980 & 0.975 & 1 & 0.535 \\
  & 2 & 1 & 1 & 1 & 0.825 \\
  & 3 & 1 & 1 & 1 & 0.965 \\
  \hline
  $p = 100, n = 20, \Sigma = \mathbf{I}_p$ & 1 & 0.630 & 0.630 & 0.560 & 0.440 \\
  & 2 & 0.915 & 0.920 & 0.700 & 0.860 \\
  & 3 & 0.955 & 0.955 & 0.710 & 0.905 \\
%  & 4 & 0.96 & 0.96 & 0.745 & 0.93 \\
%  & 5 & 0.96 & 0.965 & 0.75 & 0.94 \\
\hline
 $p = 100, n = 20, \Sigma = (0.1^{|i-j|})_{1\leq i , j \leq p}$ & 1 & 0.705 & 0.700 & 0.685 & 0.495 \\
  & 2 & 0.960 & 0.965 & 0.810 & 0.890 \\
  & 3 & 0.970 & 0.970 & 0.845 & 0.970 \\
%  & 4 & 0.99 & 0.99 & 0.615 & 0.92 \\
%  & 5 & 1 & 1 & 0.695 & 0.935 \\
  \hline
   $p = 100, n = 20, \Sigma = (0.5^{|i-j|})_{1\leq i , j \leq p}$ & 1 & 0.940 & 0.940 & 0.945 & 0.505 \\
  & 2 & 1 & 1 & 0.990 & 0.940 \\
  & 3 & 1 & 1 & 0.995 & 0.975 \\
%  & 4 & 0.99 & 0.99 & 0.615 & 0.92 \\
%  & 5 & 1 & 1 & 0.695 & 0.935 \\
\hline
$p = 100, n = 20, \Sigma = (0.9^{|i-j|})_{1\leq i , j \leq p}$ & 1 & 0.745 & 0.74 & 1 & 0.465 \\
  & 2 & 0.995 & 0.995 & 1 & 0.635 \\
  & 3 & 1 & 1 & 1 & 0.805 \\
%  & 4 & 0.99 & 0.99 & 0.615 & 0.92 \\
%  & 5 & 1 & 1 & 0.695 & 0.935 \\
   \hline
   \hline
\end{tabular}
   \caption{Proportion of times SIS, ISIS and our method selected a set of covariates containing \{$X_1,X_2,X_3$\}.}
   \label{tab:var_screening_new}
\end{table}

%\begin{figure}[htbp]
%  \includegraphics[scale = 0.36]{IR_screen.pdf}
  %\includegraphics[scale = 0.36]{IR_s_3.pdf}
%  \centering

%  \caption{\small Above: The inclusion rate for Case I with different thresholds. The green dashed line is the ISIS, where it failed when $\beta > 4$}
%  \label{fig:IR_screen}
%\end{figure}

\subsection{Study of power and size of LOCO path tests of hypotheses}
\subsubsection{Test involving a single coefficient}
We first study the size and power of the LOCO path test for testing the hypotheses $H_0$: $\beta_j = 0$ versus $H_1$: $\beta_j \neq 0$ for some $j \in \{1,\dots,p\}$, where the rejection region of the test is calibrated using the residual bootstrap procedure described in Section \ref{sec:hypothesistesting}. We consider the test statistics $T_j(1,1)$, $T_j(2,2)$, and $T_j(\infty,\infty)$. In high-dimensional ($p\geq n$) settings, we compare the empirical size and power of our test based on these statistics with the test based on the desparsified LASSO estimator of \cite{van2014asymptotically}.  We use the R package \texttt{hdi} \cite{hdi_r} to obtain the P-value based on the desparsified LASSO estimator using default settings \cite{dezeure2015high}. And we utilize the R package \texttt{lars} \cite{lars_r} with lasso modification to implement our method. In low-dimensional ($p<n$) settings, we compare the performance of our tests to that of the classical $t$-test.
\par 
We generate data according to the model

\begin{equation*}
Y = \mathbf{X}\beta + \epsilon,
\end{equation*}
where $\epsilon \sim \mathcal{N}(0, \mathbf{I}_n)$ and consider three cases with $n = 100$, $p = 80$ and $p = 1000$. For $p=1000$, we set  $\beta = (\beta_1,\dots,\beta_p)^T$ such that $\beta_2 = \dots = \beta_{10} = 1$, $\beta_{11} = \dots = \beta_{1000} = 0$. For $p=80$, we set $\beta_2 = \beta_{3} = 1$, $\beta_{4} = \dots = \beta_{80} = 0$. To simulate the power curve, we take different values of $\beta_1 \in \{0/10,1/10,\dots,1\}$.  Each row of $\mathbf{X}$ is generated independently from the multivariate normal distribution $\mathcal{N}(0,\Sigma)$, where we consider different choices of the $p \times p$ covariance matrix $\Sigma$. For each choice of $\Sigma$ and for each value of $\beta_1 \in \{0/10,1/10,\dots,1\}$, we generate $N =500$ data sets and with each data set we test $H_0$: $\beta_1 = 0$ versus $H_1$: $\beta_1 \neq 0$. For each data set, we draw $B = 500$ bootstrap samples to estimate the null distribution. We record the proportion of rejections of $H_0$ at the $\alpha = 0.05$ significance level.
\par
The empirical size of the simulation for $H_0$: $\beta_1=0$ under $p=1000$, is given in Table \ref{tab:size_all} under different choices of $\Sigma$. 
We also recorded the empirical size of the test based on the desparsified LASSO estimator. It is clear that our method nicely controlled the size under different choices of $\Sigma$ and different quantities $T_1(1,1)$, $T_1(2,2)$ and $T_1(\infty, \infty)$. The desparsified LASSO does not control the size in many cases.

\par
\begin{figure}[htbp]
  \includegraphics[scale = 0.7]{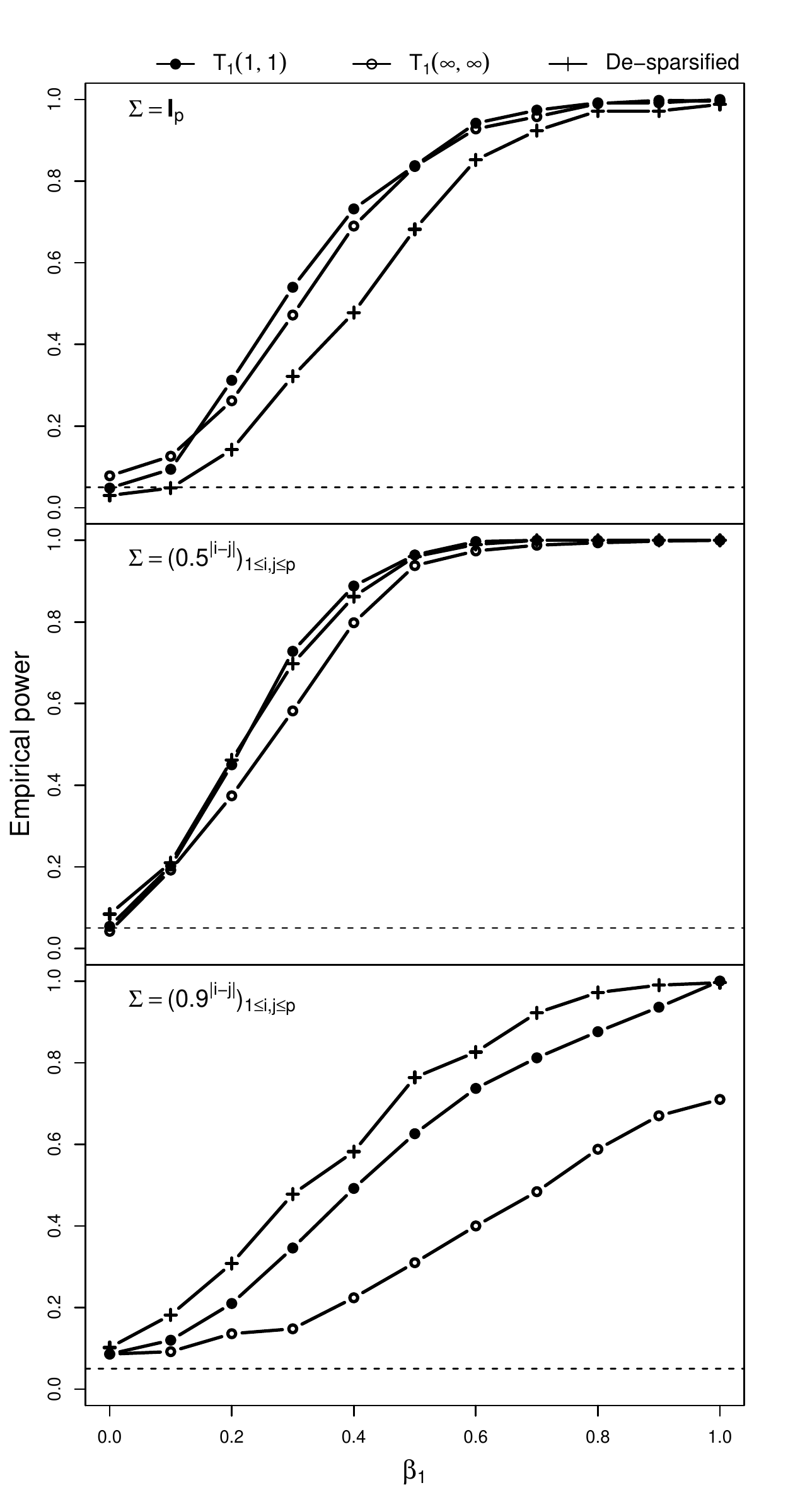}
  \centering
  \caption{\small Empirical power for testing $H_0$: $\beta_1 = 0$ vs $H_1$: $\beta_1 \neq 0$ under different correlation design with $n=100$, $p = 1000$ (from top to bottom: $\Sigma = \mathbf{I}_p$, $\Sigma = (0.5^{|i-j|})_{1\leq i , j \leq p}$, and $\Sigma = (0.9^{|i-j|})_{1\leq i , j \leq p}$). }
  \label{fig:power1000}
\end{figure}
\begin{figure}[htbp]
  \includegraphics[scale = 0.7]{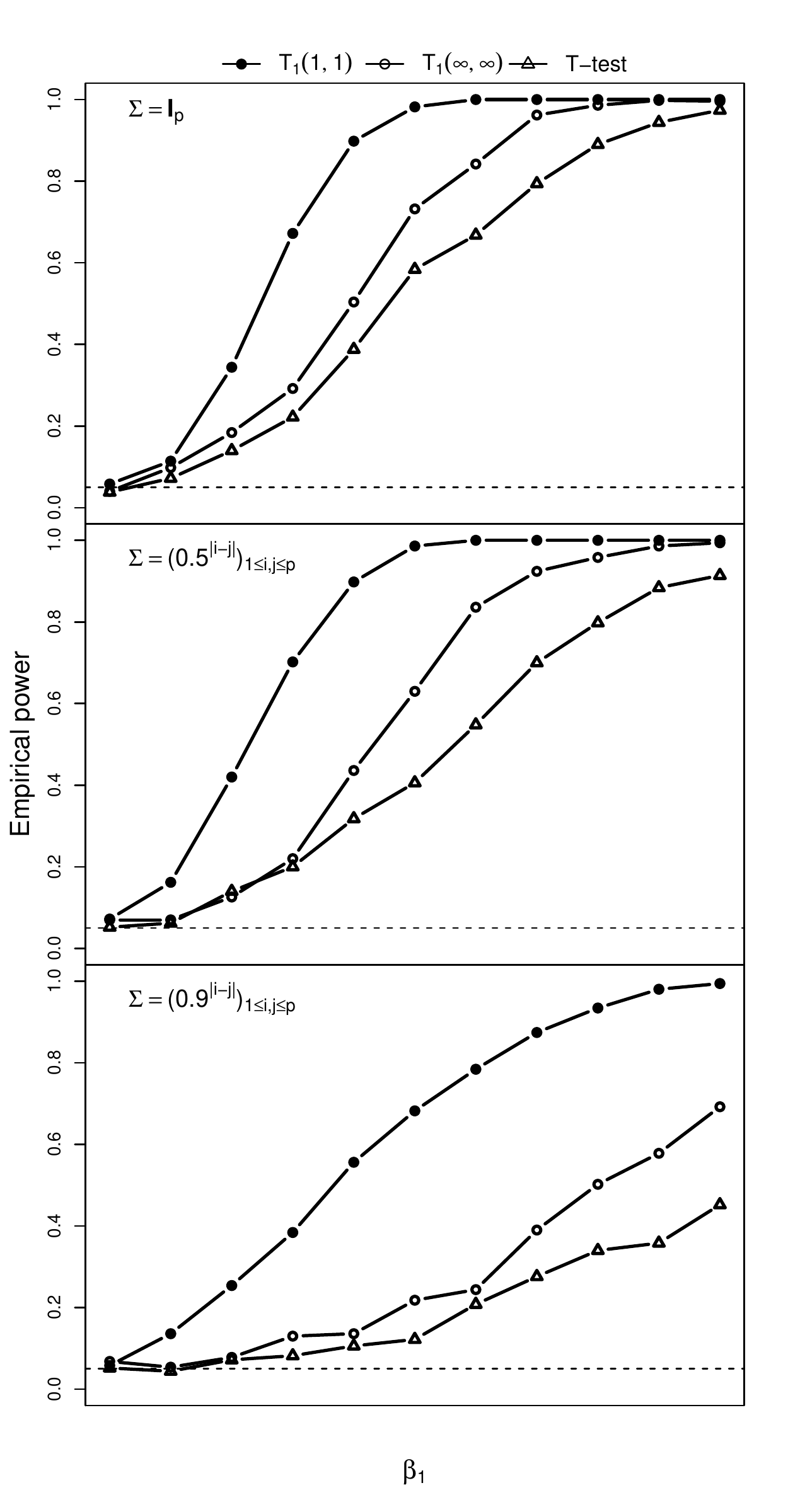}
  \centering
  \caption{\small Empirical power for testing $H_0$: $\beta_1 = 0$ vs $H_1$: $\beta_1 \neq 0$ under different correlation design with $n=100$, $p = 80$ (from top to bottom: $\Sigma = \mathbf{I}_p$, $\Sigma = (0.5^{|i-j|})_{1\leq i , j \leq p}$, and $\Sigma = (0.9^{|i-j|})_{1\leq i , j \leq p}$). }
  \label{fig:power80}
\end{figure}

The empirical power curves of our test based on the LOCO path statistics $T_1(1,1)$ and $T_1(\infty,\infty)$ as well as of the test based on the desparsified LASSO under settings $n=1000$ and $p = 80$ over the values $\beta_1 \in \{0/10,1/10,\dots,1\}$ are depicted in Figures \ref{fig:power1000} and \ref{fig:power80}. For most cases, $T_1(1,1)$ have the highest power, while $T_1(\infty,\infty )$ loses a lot of power under the correlated design. Under different designs, our method outperformed desparsified LASSO using quantity $T_1(1,1)$. It is interesting to see that the desparsified LASSO appears to outperform our method under the design $\Sigma = (0.9^{|i-j|})_{1\leq i , j \leq p}$. However, since its size is inflated in that case, we dismiss its power curve. Overall, our methods achieves comparable or higher power, with size well-controlled, compared to the desparsified LASSO method.

For the $p=80$ case, we will compare our method to the classical $t$-test. From the power curve in Figure \ref{fig:power80}, it is clear that our method achieved considerably greater power than the $t$-test using both $T_1(1,1)$ and $T_1(\infty,\infty)$, while controlling the size at the same time.

\begin{table}[htbp]
\centering
\begin{tabular}{llllll}
  \hline
  \hline
 Design & Method & $\alpha=0.20$ & $\alpha=0.10$ & $\alpha=0.05$& $\alpha=0.01$\\
  \hline
  \hline
  $\Sigma = \mathbf{I}_p$ & $T_1(1,1)$ & 0.194 & 0.106 & 0.048 & 0.008 \\
  & $T_1(2,2)$ & 0.186 & 0.110 & 0.056&  0.016 \\
  & $T_1(\infty,\infty)$ & 0.230 & 0.140 & 0.078 & 0.012  \\
  &  Desparsified  & 0.138 & 0.058 & 0.030 & 0.010 \\
  \hline
  $\Sigma = (0.5^{|i-j|})_{1\leq i , j \leq p}$&  $T_1(1,1)$ & 0.226 & 0.110 & 0.054 & 0.018 \\
  & $T_1(2,2)$ & 0.192 & 0.084 & 0.040 & 0.004 \\
  & $T_1(\infty,\infty)$ & 0.196 & 0.090 & 0.042 & 0.008  \\
  & Desparsified  & 0.222 & 0.138 & 0.084 & 0.020 \\
  \hline
  $\Sigma = (0.9^{|i-j|})_{1\leq i , j \leq p}$ & $T_1(1,1)$ & 0.214 & 0.116 & 0.086 & 0.024  \\
  & $T_1(2,2)$ & 0.238 & 0.124 & 0.076 & 0.030 \\
  & $T_1(\infty,\infty)$ & 0.264 & 0.160 & 0.086 & 0.018  \\
  & Desparsified  & 0.274 & 0.162 & 0.102 & 0.054 \\
  \hline
    $\Sigma = (0.5^{\mathbf{1}(i \neq j)})_{1\leq i , j \leq p}$ & $T_1(1,1)$ & 0.194 & 0.126 & 0.064 &  0.018  \\
  & $T_1(2,2)$ & 0.212 & 0.098 & 0.050 & 0.008 \\
  & $T_1(\infty,\infty)$ & 0.180 & 0.102 & 0.050 & 0.014  \\
  & Desparsified  & 0.126 & 0.048 & 0.028 & 0.004 \\
  \hline
   $\Sigma = (0.8^{\mathbf{1}(i \neq j)})_{1\leq i , j \leq p}$ & $T_1(1,1)$ & 0.242 & 0.116 & 0.056 & 0.010 \\
  & $T_1(2,2)$ & 0.182 & 0.086 & 0.040 & 0.010 \\
  & $T_1(\infty,\infty)$ & 0.198 & 0.084 & 0.050 & 0.008   \\
  & Desparsified  & 0.070 & 0.022 & 0.010 & 0.002 \\
   \hline
   \hline
\end{tabular}
   \caption{Empirical size of the test under different $\Sigma$ with $n=100$, $p=1000$.}
   \label{tab:size_all}
\end{table}

\par

\subsubsection{Test involving multiple coefficients}
For the simultaneous test, we consider similar settings. For $p=1000$, We will test 
\[
\text{$H_0$: $\beta_1 = 1$, $\beta_{11} = 0$, $\beta_{12} = 0$ vs $H_1$: $\beta_1 \neq 1$ or  $\beta_{11} \neq 0$ or $\beta_{12} \neq 0$.}
\]
and for $p=80$, we will test 
\[
\text{$H_0$: $\beta_1 = 1$, $\beta_{4} = 0$, $\beta_{5} = 0$ vs $H_1$: $\beta_1 \neq 1$ or  $\beta_{4} \neq 0$ or $\beta_{5} \neq 0$.}
\]

We generate data according to the model

\begin{equation*}
Y = \mathbf{X}\beta + \epsilon,
\end{equation*}
where $\epsilon \sim \mathcal{N}(0, \mathbf{I}_n)$ with $n = 100$ and $\beta = (\beta_1,\dots,\beta_p)^T$. For $p=1000$, we set $\beta_2 = \dots = \beta_{10} = 1$, $\beta_{11} = \dots = \beta_{1000} = 0$, and $\beta_1 \in \{1,11/10,\dots,2\}$. For $p=80$, we set $\beta_2 = \beta_{3} = 1$. Other settings remain the same as those under which we tested $H_0$: $\beta_1 = 0$ versus $H_1$: $\beta_1 \neq 0$. 

%We compared our results to \cite{zhang2017simultaneous}.

\begin{figure}[htbp]
  \includegraphics[scale = 0.7]{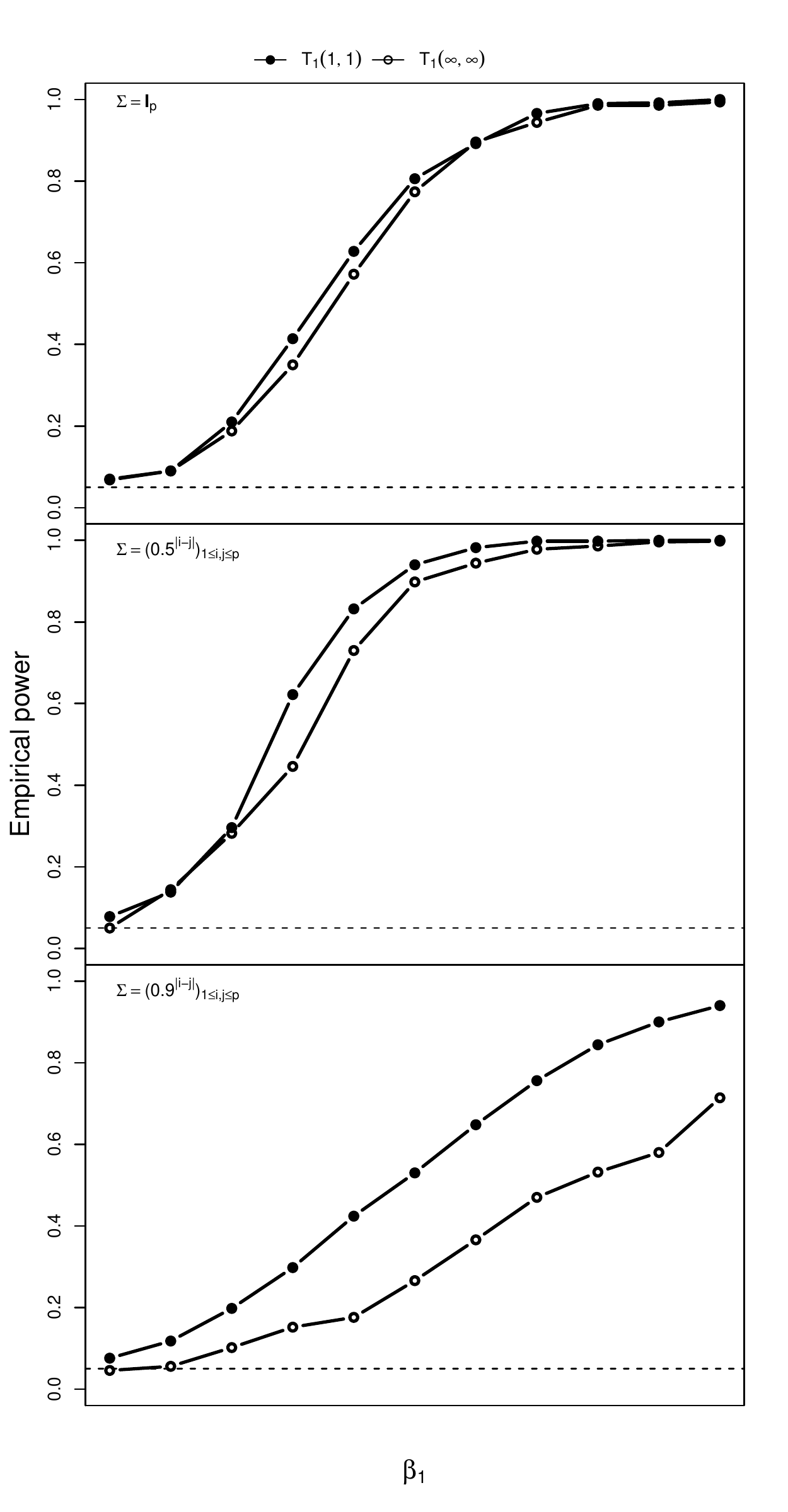}
  \centering
  \caption{\small Multiple testing empirical power under different correlation design with $n=100$, $p = 1000$ (from top to bottom: $\Sigma = \mathbf{I}_p$, $\Sigma = (0.5^{|i-j|})_{1\leq i , j \leq p}$, and $\Sigma = (0.9^{|i-j|})_{1\leq i , j \leq p}$). }
  \label{fig:powermulti}
\end{figure}

For the $p=1000$ case, Figure \ref{fig:powermulti} shows the power curves of the tests under different choices of $\Sigma$. The size is well controlled when $H_0$ is true, and $T_1(1,1)$ achieved higher power than $T_1(\infty,\infty)$ as the correlation increases.. 

\begin{figure}[htbp]
  \includegraphics[scale = 0.7]{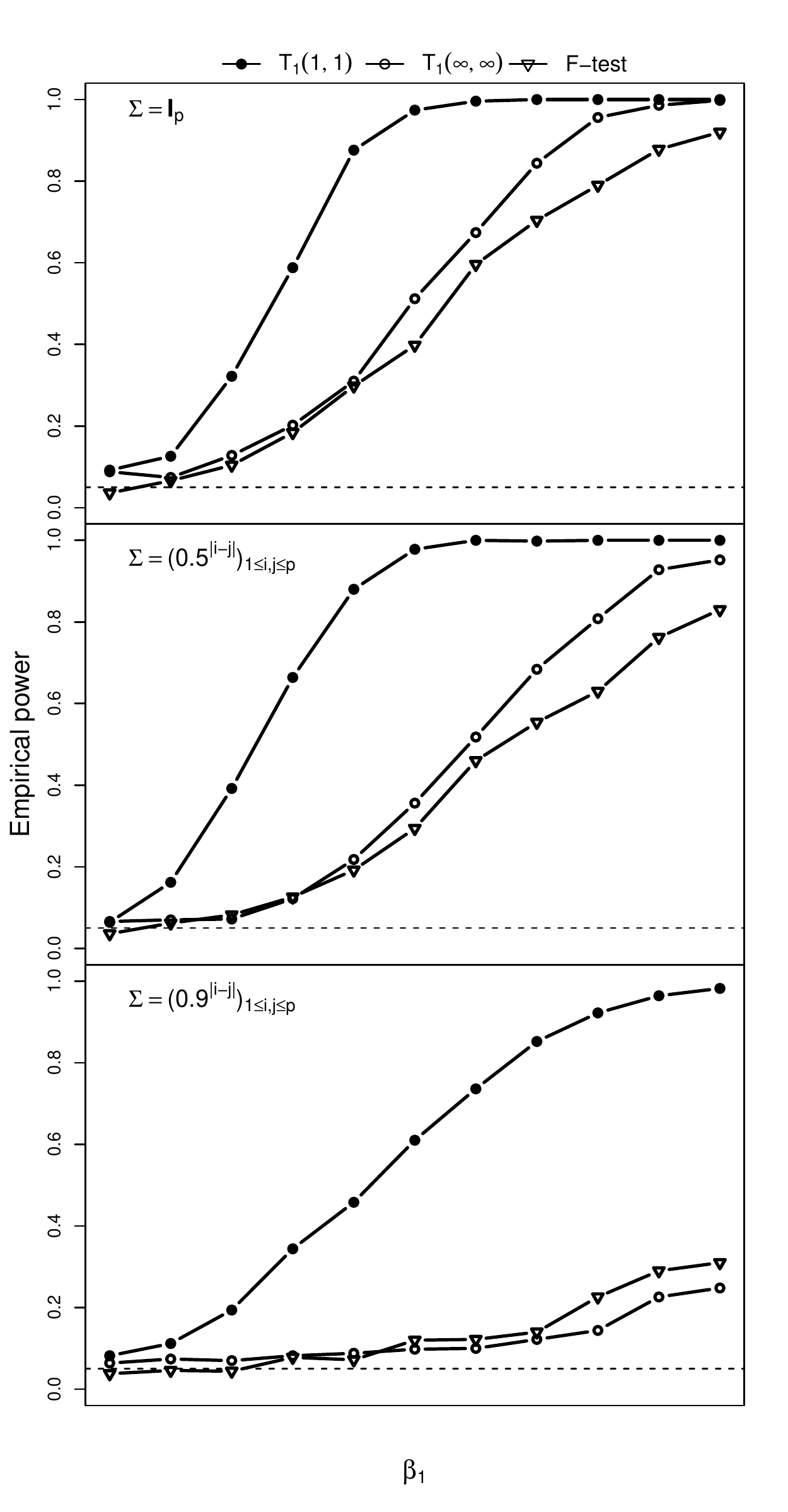}
  \centering
  \caption{\small Multiple testing empirical power under different correlation design with $n=100$, $p = 80$ (from top to bottom: $\Sigma = \mathbf{I}_p$, $\Sigma = (0.5^{|i-j|})_{1\leq i , j \leq p}$, and $\Sigma = (0.9^{|i-j|})_{1\leq i , j \leq p}$). }
  \label{fig:multi_power_80}
\end{figure}

For the $p=80$ case, we will compare our method to the classical F-test. From the power curve in Figure \ref{fig:multi_power_80}, it is clear our method achieved considerably greater power than the F-test both $T_1(1,1)$ and $T_1(\infty,\infty)$, while controlling the size at the same time.

% ST and NST are different methods in \cite{zhang2017simultaneous}.   % Results under the other choices of $\Sigma$ are presented in the Supplementary Material.

\section{Real data analysis}\label{sec:dataanalysis}

To provide a concrete example, we consider a dataset about riboflavin (vitamin B2) production in Bacillus subtilis with 71 observations and 4088 variables \cite{buhlmann2014high} \cite{dezeure2015high} \cite{van2014asymptotically}. The response variable measures the logarithm of the riboflavin production rate and the predictors are logarithm of the expression level of 4088 genes. We will model the data with a high-dimensional linear model and carry out variable screening and inferences with the LOCO path statistic.
\par
\begin{figure}[htbp]
\includegraphics[width=.8\textwidth]{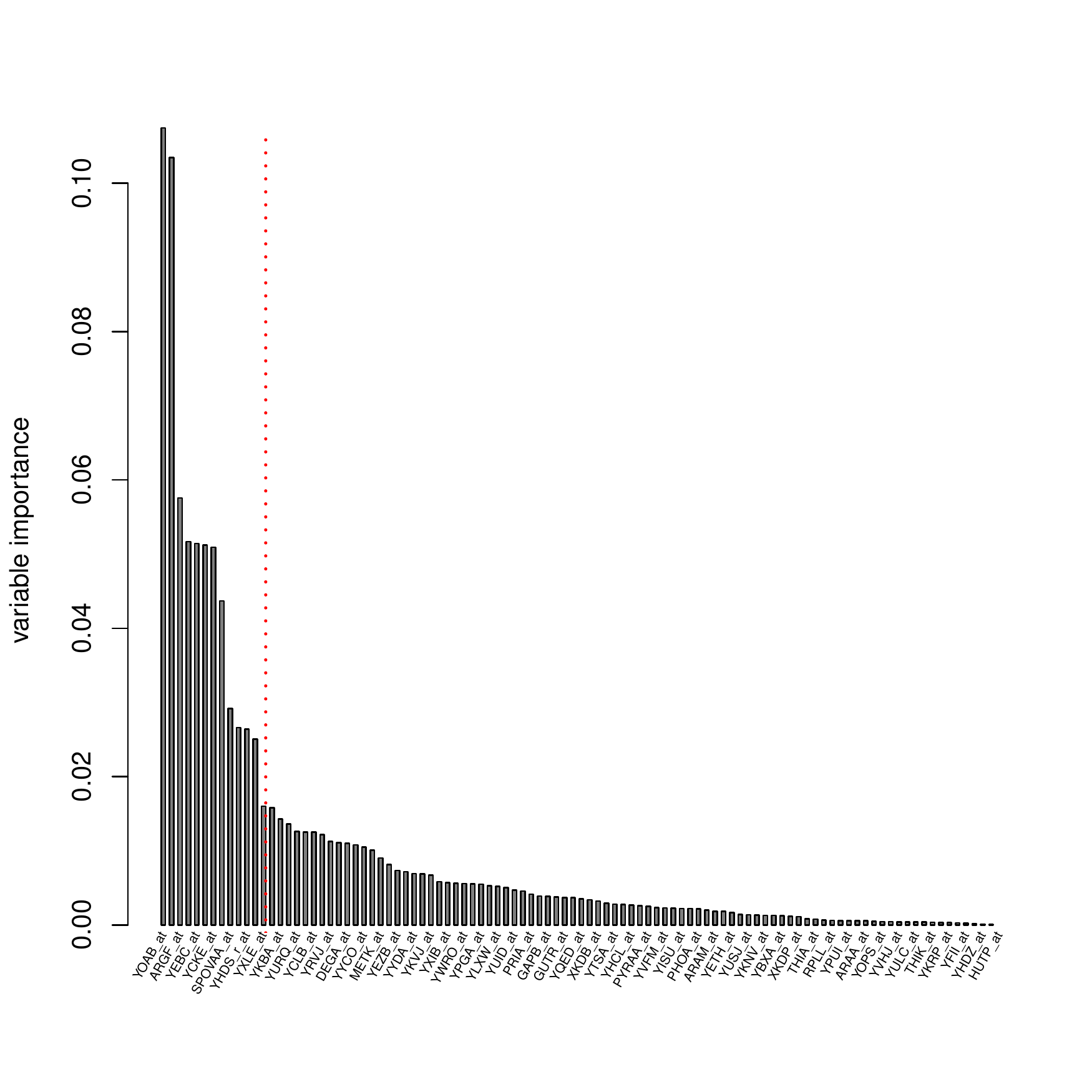}
  \includegraphics[width=.8\textwidth]{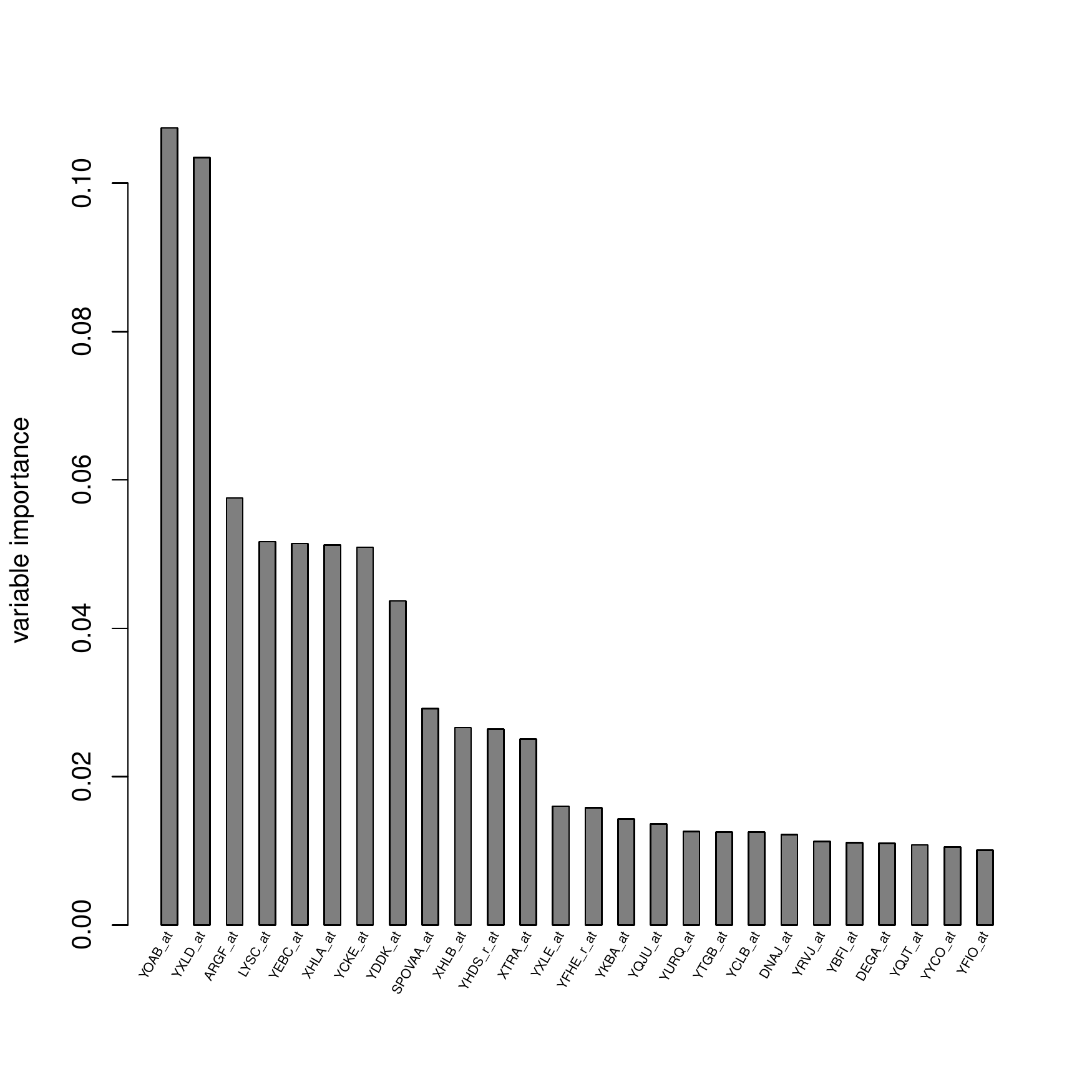}
  \centering
  \caption{\small Top: The first 100 most important genes. The vertical dotted line marks the variable importance at 1\%. Bottom: All genes with variable importance $> 1\%$. }
  \label{fig:ts_ribo}
\end{figure}

We use $T(1,1)$ in this part and obtained bootstrap P-values for each gene after variable screening. We screened in 342 genes with $T_j(1,1) > 0$, $j = 1,\dots, 4088$. Based on our bootstrapped P-values, our method found the following 9 significant genes at 0.05 significance level: ARGF\_at, XHLA\_at, XHLB\_at, XTRA\_at, YCKE\_at, YEBC\_at, YOAB\_at,  YXLD\_at and YYBG\_at. Using the P-values based on the desparsified LASSO results in 0 significant genes \cite{dezeure2015high}. Figure \ref{fig:ts_ribo} shows the variable importance for a small portion of genes. We will see only a few genes have large variable importance, while most genes have variable importance  less than 1\%.

% 
%After applying Benjamini–Hochberg procedure, 
%only 3 genes were selected XHLA\_at, %YOAB\_at and YXLD\_at.
%\bl{[Explain which test statistic you used.  Was is $T_j(1,1)$?  Explain how you got P-values for all of the genes using the proposed method.]}.

Table \ref{tab:var_imp} shows all variables with importance $1\%$,  where YXLD\_at and YOAB\_at have the largest variable importance. Both genes are also tested significant using our bootstrap procedure.

%
%[1] "YOAB_at"   "YXLD_at"   "LYSC_at"   "XHLA_at"   "YEBC_at"   "YCKE_at"   "YDDK_at"   "ARGF_at"   "SPOVAA_at"
%[10] "XHLB_at"  
%> TS_perct[sig_index]
 %[1] 0.10951092 0.10531367 0.05314275 0.05255299 0.05252845 0.05223592 0.04041187 %0.03770742 0.03100400 0.02734934
%
%colnames(riboflavin$x)[sig_index]
% "ARGF_at"   "SPOIVA_at" "THIA_at"   "XHLA_at"   "XHLB_at"   "XTRA_at"
% "YCKE_at"   "YDDK_at"   "YEBC_at"   "YOAB_at"   "YRVJ_at"   "YURQ_at"
%"YUSJ_at"   "YXCA_at"   "YXLD_at"   "YYBG_at"

%> pval[which(pval<0.05)]
% [1] 0.00 0.00 0.00 0.00 0.00 0.00 0.00 0.00 0.00 0.00 0.00 0.04 0.04 0.00 0.00
%[16] 0.00

%

\begin{table}[htbp]
\centering
\begin{tabular}{rll}
  \hline
 Genes & Importance & P-value\\
  \hline
 YOAB\_at & 10.7\% & 0.0084\\
   YXLD\_at & 10.3\% & 0.0084\\
   ARGF\_at & 5.8\% & 0.0168\\
   LYSC\_at & 5.2\% & 0.0924\\
   YEBC\_at & 5.2\% & 0.0616\\
   XHLA\_at & 5.1\% & 0.0140\\
   YCKE\_at & 5.1\% & 0.0084\\
   YDDK\_at & 4.4\% & 0.0560\\
   SPOVAA\_at & 2.9\% & 0.1482\\
   XHLB\_at & 2.7\%  & 0.0194\\
   \hline
\end{tabular}
   \caption{The first 10 most important genes.}
    \label{tab:var_imp}
\end{table}

\section{Discussion}\label{sec:discussion}

Our LOCO path statistic provides a new way to do variable screening and statistical inference in linear models. For variable screening, our method does not require the selection of tuning parameters and can achieve a greater probability of selecting a set of covariates that contains the true model than both SIS and ISIS.
For statistical inference, our method provides reliable P-values in both high and low-dimensional settings. Overall, the proposed bootstrap method controls the size and in some cases achieves higher power than the desparsified LASSO of \cite{van2014asymptotically}. Moreover, our method can be used to test hypothesis simultaneously involving multiple coefficients.
We believe the LOCO path idea can be readily extended to other settings.
\par
Consider the regularization optimization problem
\begin{equation}\label{eqn:minimizationproblem}
    \hat{\beta} = \hat{\beta}(\lambda) \defeq \operatorname*{argmin}_{\beta \in \mathbb{R}^p} L(Y,  \mathbf{X}\beta)  + \lambda J(\beta),
\end{equation}
where $L(\cdot)$ is a pre-defined loss function, $\lambda > 0$ is a tuning parameter which controls the level of regularization, and $J(\cdot)$ is a penalty function on $\beta$. The solution path $\hat{\beta}(\lambda)$ could be viewed as a $1$-to-$p$ mapping $\lambda \mapsto \hat{\beta}(\lambda)$ taking values in $(0,\infty)$ and returning values in $\mathbb{R}^p$. Since our measure of feature importance and variable screening procedure relies on the solution path only, we can easily adapt our method to \eqref{eqn:minimizationproblem}, which includes logistic regression, Poisson regression and Cox models. Appropriate bootstrap methods for calibrating hypothesis tests would have to be worked out under each setting, which we leave to future work.

\section*{Acknowledgements}{This work was partially supported by Grant R03 AI135614 from the National Institutes of Health.}

\section*{Supplementary Material}{Supplementary material related to this article can be found in our submission.} 

\bibliographystyle{plain}
\bibliography{pp}

\begin{thebibliography}{10}

\bibitem{breiman2001random}
Leo Breiman.
\newblock Random forests.
\newblock {\em Machine learning}, 45(1):5--32, 2001.

\bibitem{buhlmann2014high}
Peter B{\"u}hlmann, Markus Kalisch, and Lukas Meier.
\newblock High-dimensional statistics with a view toward applications in
  biology.
\newblock {\em Computational Statistics}, 29:407--430, 2014.

\bibitem{chatterjee2013rates}
Arindam Chatterjee, Soumendra~N Lahiri, et~al.
\newblock Rates of convergence of the adaptive lasso estimators to the oracle
  distribution and higher order refinements by the bootstrap.
\newblock {\em The Annals of Statistics}, 41(3):1232--1259, 2013.

\bibitem{chatterjee2011bootstrapping}
Arindam Chatterjee and Soumendra~Nath Lahiri.
\newblock Bootstrapping lasso estimators.
\newblock {\em Journal of the American Statistical Association},
  106(494):608--625, 2011.

\bibitem{das2019perturbation}
Debraj Das, Karl Gregory, SN~Lahiri, et~al.
\newblock Perturbation bootstrap in adaptive lasso.
\newblock {\em The Annals of Statistics}, 47(4):2080--2116, 2019.

\bibitem{hdi_r}
Ruben Dezeure, Peter B\"uhlmann, Lukas Meier, and Nicolai Meinshausen.
\newblock High-dimensional inference: Confidence intervals, p-values and
  {R}-software {hdi}.
\newblock {\em Statistical Science}, 30(4):533--558, 2015.

\bibitem{dezeure2015high}
Ruben Dezeure, Peter B{\"u}hlmann, Lukas Meier, Nicolai Meinshausen, et~al.
\newblock High-dimensional inference: Confidence intervals, $ p $-values and
  r-software hdi.
\newblock {\em Statistical Science}, 30(4):533--558, 2015.

\bibitem{fan2011nonparametric}
Jianqing Fan, Yang Feng, and Rui Song.
\newblock Nonparametric independence screening in sparse ultra-high-dimensional
  additive models.
\newblock {\em Journal of the American Statistical Association},
  106(494):544--557, 2011.

\bibitem{fan2008sure}
Jianqing Fan and Jinchi Lv.
\newblock Sure independence screening for ultrahigh dimensional feature space.
\newblock {\em Journal of the Royal Statistical Society: Series B (Statistical
  Methodology)}, 70(5):849--911, 2008.

\bibitem{fan2010sure}
Jianqing Fan, Rui Song, et~al.
\newblock Sure independence screening in generalized linear models with
  np-dimensionality.
\newblock {\em The Annals of Statistics}, 38(6):3567--3604, 2010.

\bibitem{fisher2018all}
Aaron Fisher, Cynthia Rudin, and Francesca Dominici.
\newblock All models are wrong but many are useful: Variable importance for
  black-box, proprietary, or misspecified prediction models, using model class
  reliance.
\newblock {\em arXiv preprint arXiv:1801.01489}, 2018.

\bibitem{lars_r}
Trevor Hastie and Brad Efron.
\newblock {\em lars: Least Angle Regression, Lasso and Forward Stagewise},
  2013.
\newblock R package version 1.2.

\bibitem{javanmard2014confidence}
Adel Javanmard and Andrea Montanari.
\newblock Confidence intervals and hypothesis testing for high-dimensional
  regression.
\newblock {\em The Journal of Machine Learning Research}, 15(1):2869--2909,
  2014.

\bibitem{ke2014covariance}
Tracy Ke, Jiashun Jin, and Jianqing Fan.
\newblock Covariance assisted screening and estimation.
\newblock {\em Annals of Statistics}, 42(6):2202--2242, 2014.

\bibitem{lei2018distribution}
Jing Lei, Max G’Sell, Alessandro Rinaldo, Ryan~J Tibshirani, and Larry
  Wasserman.
\newblock Distribution-free predictive inference for regression.
\newblock {\em Journal of the American Statistical Association},
  113(523):1094--1111, 2018.

\bibitem{lockhart2014significance}
Richard Lockhart, Jonathan Taylor, Ryan~J Tibshirani, and Robert Tibshirani.
\newblock A significance test for the lasso.
\newblock {\em Annals of Statistics}, 42(2):413--468, 2014.

\bibitem{mammen2012does}
Enno Mammen.
\newblock {\em When does bootstrap work?: asymptotic results and simulations},
  volume~77.
\newblock Springer Science \& Business Media, 2012.

\bibitem{meinshausen2009p}
Nicolai Meinshausen, Lukas Meier, and Peter B{\"u}hlmann.
\newblock P-values for high-dimensional regression.
\newblock {\em Journal of the American Statistical Association},
  104(488):1671--1681, 2009.

\bibitem{SIS_r}
Diego~Franco Saldana and Yang Feng.
\newblock {SIS}: An {R} package for sure independence screening in
  ultrahigh-dimensional statistical models.
\newblock {\em Journal of Statistical Software}, 83(2):1--25, 2018.

\bibitem{tibshirani1996regression}
Robert Tibshirani.
\newblock Regression shrinkage and selection via the lasso.
\newblock {\em Journal of the Royal Statistical Society. Series B
  (Methodological)}, 58(1):267--288, 1996.

\bibitem{van2014asymptotically}
Sara Van~de Geer, Peter B{\"u}hlmann, Ya’acov Ritov, Ruben Dezeure, et~al.
\newblock On asymptotically optimal confidence regions and tests for
  high-dimensional models.
\newblock {\em The Annals of Statistics}, 42(3):1166--1202, 2014.

\bibitem{wasserman2009high}
Larry Wasserman and Kathryn Roeder.
\newblock High dimensional variable selection.
\newblock {\em Annals of Statistics}, 37(5A):2178--2201, 2009.

\bibitem{zhang2014confidence}
Cun-Hui Zhang and Stephanie~S Zhang.
\newblock Confidence intervals for low dimensional parameters in high
  dimensional linear models.
\newblock {\em Journal of the Royal Statistical Society: Series B (Statistical
  Methodology)}, 76(1):217--242, 2014.

\bibitem{zou2006adaptive}
Hui Zou.
\newblock The adaptive lasso and its oracle properties.
\newblock {\em Journal of the American Statistical Association},
  101(476):1418--1429, 2006.

\end{thebibliography}

\end{document}